\documentstyle[prl,aps,epsf]{revtex} 
\newcommand{\di}{{d}}   

\newcommand{\Vec}[1]{\mbox{\boldmath$#1$}} 
\newcommand{\unitVec}[1]{\hat{\mbox{\boldmath$#1$}}}

\begin{document} 

\title{The effect of an external magnetic field on the gas-liquid 
transition in the Heisenberg spin fluid}

\author{T.G.Sokolovska, R.O.Sokolovskii\cite{email}}
\address{Institute for Condensed Matter Physics,
Svientsitskii 1, Lviv 290011, Ukraine. }

\maketitle 

\begin{abstract}
We present the theoretical phase diagrams of the classical Heisenberg 
fluid in an external magnetic field. A consistent account of correlations 
is carried out by the integral equation method. A nonmonotoneous 
effect of fields on the temperature of the gas-liquid critical point is 
found. Within the mean spherical approximation this nonmonotoneous 
behavior disappears for short-range enough spin-spin interactions.
\end{abstract}

\begin{center}
PACS numbers 61.20.Gy, 64.70.Fx, 75.50.Mm, 05.70.Jk.
\end{center}
%\pacs{61.20.Gy, 64.70.Fx, 75.50.Mm, 05.70.Jk}

The behavior of magnetic fluids in an external field  
compels more and more attention for the last years and
has some peculiarities. 
In 
the presence of an external magnetic field the orientational (magnetic) 
phase transition is absent, but there are the first order transitions 
between ferromagnetic phases of different densities (e.g., gaseous and 
liquid phases). Physical properties of anisotropic fluids (to these belong 
also, besides magnetic fluids, nematic liquid crystals) are determined by 
the interplay between orientational and translational degrees of freedom.  
Therefore, varying the magnetic field it is possible to effect structural 
properties of magnetic fluids, in particular, to change the region of the 
gas-liquid coexistence.  Such investigations with a calculation of phase 
diagrams were carried out for model spin systems within the mean field 
(MF) approximation \cite{Kawasaki,Schinagl}. It was found that for fluids 
of hard spheres carrying Ising spins an external magnetic field decreases 
the temperature of the gas-liquid critical point. On the other hand, the 
presence of isotropic van~der~Waals attractions between molecules can lead 
to the inverse effect \cite{Schinagl}.  
In Ref.~\cite{Schinagl} 
the fluid of hard spheres with the classical Heisenberg spins and strong 
isotropic attractions was considered also. It was shown that at weak 
magnetic fields there can be two first order phase transitions in this 
model:  gas-liquid and liquid-liquid.  In strong fields the weak 
liquid-liquid transition disappears.

The need to account orientational-translational correlations for the 
description of physical properties of magnetic fluids stimulated studies 
of the external field effects by more complex techniques. The 
effect of an external field on the gas-liquid critical point was 
studied by the functional integration and Green function methods 
\cite{VRP} for the quantum Heisenberg ferrofluid and by the Monte-Carlo 
and integral equation methods for the classical one 
\cite{Lado,LadoPRL}.  The pair potentials of those models consisted of 
contributions of hard spheres and of the spin-spin interaction (the 
so-called ideal Heisenberg fluid).  In these works the conclusion was made 
that an external magnetic field favors the phase separation, i.e., the 
application of the external field increases the gas-liquid critical 
temperature. Let us note that the results of Refs.~\cite{Lado,LadoPRL} are 
obtained for quite strong fields.  In our point of view, it was the effect 
of small fields that is of special interest.  This follows from the fact 
that at small fields orientational fluctuations are large and the 
corresponding correlations have a long-range character.  Therefore, small 
external influences can result in significant changes of macroscopic 
properties of magnetic fluids. Besides, the interplay between 
orientational and spatial ordering can lead to an interesting behavior of 
the gas-liquid critical point at small fields. In this letter we show that 
in the systems with the long-range enough spin-spin interactions the 
nonmonotoneous effect of the external fields on the gas-liquid critical 
temperature takes place. There is a temperature range, where weak external 
fields suppress the gas-liquid phase separation. By the integral equation 
method we show that while the range parameter of the model potential 
decreases this temperature interval first gets smaller and then 
disappears.

\begin{figure}
\noindent
\epsfxsize=85mm
\centerline{\epsffile{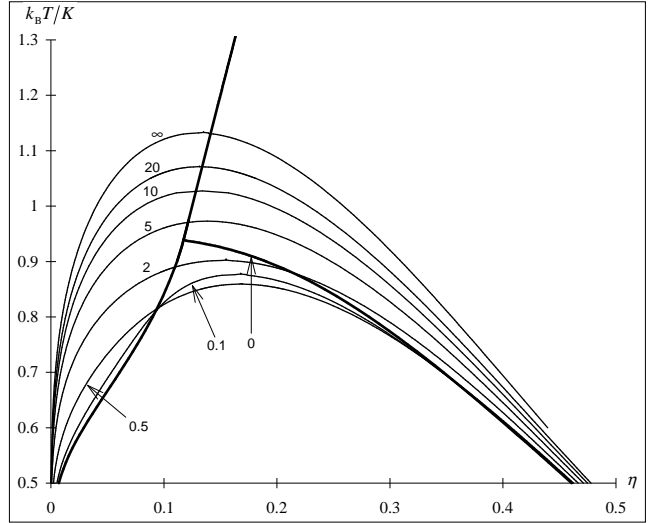}}
\caption{
Phase diagram of the Heisenberg fluid in the external magnetic field. MF 
approximation. The results are the same for any $z\sigma$. The thick
lines constitute the phase diagram (the Curie line and the gas-liquid 
binodal) for $h=0$.  The thin lines are the gas-liquid coexistence lines 
for $h=0.1$, 0.5, 2, 5, 10, 20, $\infty$ }
\label{mfa}
\end{figure}

We shall consider the model that was studied before for the case of zero 
external field \cite{Lomba,PhysA}. The pair potential (analogically to 
the papers \cite{VRP,Lado}) is a sum of the hard sphere potential 
$\varphi(R_{12})$ for spheres of diameter $\sigma$ and of the 
Heisenberg spin-spin potential $\Phi(R_{12},\omega_1,\omega_2)$
\begin{eqnarray}
&&\Phi(R_{12},\omega_1,\omega_2)=
J(R_{12})\unitVec{S}_{1}\cdot\unitVec{S}_{2}
\label{1}
,\\
&&J(R)=-K\frac{(z\sigma)^{2}}{z\sigma+1}\frac{\exp(-z(R-\sigma))}{R/\sigma}
\label{2}
,
\end{eqnarray}
where $\unitVec{S}_{i}$ is a unit vector in the direction 
$\omega_i=(\theta_i,\varphi_i)$ of the magnetic dipole moment $\Vec{\mu}$, 
referred to the uniform field $B_0$ as the $z$ direction. The potential of 
the particle interacting with the field is $v_i=-\mu B_0\cos\theta_i$. In 
expression (\ref{2}) the coefficient $\frac{(z\sigma)^{2}}{z\sigma+1}$ is 
chosen to make the integral 
\begin{equation}
I=\frac{N^2}{2V}
 \int\di\omega_1\int\di\omega_2\int\limits_{R_{12}>\sigma}\di\Vec R_{12} 
        \Phi(R_{12},\omega_1,\omega_2)f(\omega_1)f(\omega_2)
\end{equation}
(where $f(\omega)$ is a single-particle orientational distribution 
function) independent of $z\sigma$. The integral $I$ describes a 
contribution of the spin-spin potential into the free energy functional 
within the MF approximation (see, for example, 
\cite{Tavares}).  Therefore, within the MF approach the model 
phase diagram is independent of $z\sigma$, if we use dimensionless units 
for the temperature $t=k_{\rm B}T/K$, the density $\eta=\frac 
NV\frac{\pi\sigma^3}6$, and the external field strength $h=\frac{\mu 
B_0}{K\sqrt3}$.  For the free energy of the hard sphere system we use the 
``quasi-exact'' Carnahan-Starling expression \cite{Carnahan}.  The MF 
phase diagrams obtained by the well-known double-tangent construction are 
presented in Fig.~\ref{mfa}.  One can see a nonmonotoneous effect of the 
external field on the temperature of the gas-liquid critical point at 
small values of $h$.  With the increase of the external field strength the 
gas-liquid critical point first moves down ($h=0.1$, 0.5), then moves up 
($h=2$, 5), and last ($h=10$, 20, $\infty$) grows higher than the top of 
the zero field binodal.  It should be noted that the model potential 
(\ref{2}) in the limit $z\sigma\to0$ belongs to a family of the so-called 
Kac potentials, for which the MF approach is accurate. Therefore, relying 
on the MF results (Fig.~\ref{mfa}) we can state that for long-range enough 
spin-spin interactions ($z\sigma\to0$) the nonmonotoneous field effect on 
the critical temperature does take place.  But for finite values of 
$z\sigma$ we are forced to carry out more accurate investigation.

More consistent consideration of anisotropic fluids can be done on the 
basis of the integral equation method that allows to calculate both the
one-particle distribution function and the pair distribution function. The 
task consists \cite{Henderson} in a solution of the anisotropic 
Ornstein-Zernike (OZ) equation
\begin{equation}
h(1,2)=c(1,2)+\int\rho (3)h(1,3)c(3,2)d(3),~
\label{3}
\end{equation}
where $d(3)=d\Vec{R}_{3} d\omega_{3}$, $\rho(1)=\rho f(\omega_1)$, 
$h(1,2)$ and $c(1,2)$ are the total and direct correlation functions of 
the system. Since Eq.~(\ref{3}) contains the one-particle distribution 
function, we need (besides a closure relation for the anisotropic OZ 
equation) some additional relation for determination of $\rho(1)$. It can 
be the first equation of the Bogolubov-Born-Green-Kirkwood-Yvon hierarchy 
(this was used in Refs.~\cite{Lado,LadoPRL} to obtain a numerical solution 
of Eq.~(\ref{3})) or the Lovett equation for anisotropic fluids 
\cite{Lovett}
\begin{equation}
\Vec\nabla_{\omega_{1}}\ln\rho (1)+\Vec\nabla_{\omega_{1}}
\frac{v(1)}{k_{\rm B}T}=\int c(1,2)\Vec\nabla_{\omega_{2}}\rho (2)d(2)
\label{4}
,
\end{equation}
where $\Vec\nabla_{\omega_{1}}$ is an angular gradient operator; $v(1)$ 
is a potential of interaction with a uniform external field, its 
spherical harmonic expansion is of the form
\begin{equation}
v(1)=-\sum_l v_l Y_{l0}(\omega_1).
\end{equation}
For a self-consistent solution of 
Eqs.~(\ref{3}) and (\ref{4}) we shall use the 
method suggested in \cite{PhysA,UFZH}. 
The method is based on the algebraic representation of the Lovett equation 
for uniaxial fluids. Using the general expansion for the direct 
correlation function of linear molecules
\begin{equation}
c(1,2)=\sum_{mnl\atop\mu\nu\lambda}
c_{mnl}^{\mu\nu\lambda} (R)Y_{m\mu} (\omega_{1})Y_{n\nu}^{*}
(\omega_{2})Y_{l\lambda} (\omega_{R})
\label{6}
\end{equation}
and the exponential form of the one-particle distribution function
\begin{equation}
f(\omega)=Z^{-1}
\exp\left(\sum\limits_{l>0}A_{l}Y_{l0} (\omega)\right)
\label{7}
\end{equation}
we obtain following \cite{PhysA,cmp} an
algebraic representation of the Lovett equation for a uniaxial fluid in 
the external field 
\begin{equation} 
L_l=\sum_{mn} C_{lm} Y_{mn} L_n+V_l=\sum_m C_{lm} {\cal P}_m+V_l 
\label{8}, 
\end{equation} 
where all indices take the values greater or equal one, 
$V_l=v_l/k_{\rm B}T$,
$C_{mn}=\int c_{mn0}^{110} (R)d\Vec{R}$,
$Y_{mn}=\rho\left\langle
  Y_{m1}(\omega)Y_{n1}^{*} (\omega)
\right\rangle_{\omega}$,
$\left\langle\cdots\right\rangle_{\omega}
=\int f(\omega)(\cdots)d\omega$,
$L_{l}=\sqrt{l(l+1)} A_{l}$,
${\cal P}_{l}=\rho\sqrt{l(l+1)(2l+1)}
\left\langle P_{l} (\cos\theta)\right\rangle_{\omega}$, $P_l(\cos\theta)$ 
is the $l$th order Legendre polynomial.  Let us note that the average 
values $\left\langle P_{l}(\cos\theta)\right\rangle_{\omega}$ play the 
role of order parameters in anisotropic fluids. Relations (\ref{8}) are 
accurate, and their use, as well as the use of the integro-differential 
equation (\ref{4}), does not introduce any approximation into the theory.

Due to (\ref{8}) it turns to be possible to obtain for the model 
(\ref{1})--(\ref{2}) an analytical solution of the anisotropic OZ equation 
(\ref{3}) within the mean spherical closure 
\begin{equation}
\begin{array}{ll}
c(1,2)=-\Phi(R_{12},\omega_1,\omega_2)/k_{\rm B}T,
        &R_{12}>\sigma,\\
h(1,2)=-1,
        &R_{12}<\sigma.
\end{array}
\label{9}
\end{equation}
Condition (\ref{9}) for $h(1,2)$ follows directly from the fact
that hard spheres do not overlap. The mean spherical closure (\ref{9}) 
restricts correlation functions of our model to those of the form
\begin{equation}
f(1,2)=\sum_{l_1l_2m} 
        f_{l_1l_2m}(R_{12})Y_{l_1m}(\omega_1)Y^*_{l_2m}(\omega_2),~
\label{10}
\end{equation}
($l_1,~l_2=0,1$), and representation (\ref{8}) results in equalities
\begin{eqnarray}
A_1&=&A_1\rho\left\langle\left| Y_{11} (\omega)\right|^{2}
        \right\rangle_{\omega}\int c_{111}(R)d\Vec{R}
 +\frac{v_1}{k_{\rm B}T}
\nonumber,
\\
A_1&=&A_1\rho
 \left\langle Y_{10} (\omega)\right\rangle_{\omega}\int c_{111}(R)d\Vec{R}
 +\frac{v_1}{k_{\rm B}T}
\label{11}.
\end{eqnarray}
Here we use the notations of Eq.~(\ref{10}) for harmonics of the direct 
correlation function $c(1,2)$; $v_1=\mu B_0/\sqrt{3}$. Thus, the use of 
the mean spherical closure yields for our model vanishing of coefficients 
$A_l$ with $l>1$ in Eq.~(\ref{7}), and the self-consistent one-particle 
distribution function in the mean spherical approximation
(MSA) takes the form
\begin{equation}
f(\omega)=\exp\left(A_1 Y_{10} (\omega)\right)
        /\int\exp\left(A_1 Y_{10}(\omega)\right)\di\omega 
\label{12}.
\end{equation}
A uniaxial symmetry of our system leads to factorization of Eq.~(\ref{3}) 
on the equations with different $m$. At $m=\pm1$
\begin{eqnarray}
h_{11m} (R_{12})&=&c_{11m} (R_{12})
\label{14}\\ \nonumber
&+&\rho
\left\langle\left| Y_{1m} (\omega)\right|^{2}\right\rangle_{\omega}
\int c_{11m} (R_{13})h_{11m} (R_{32}) \di\Vec{R}_{3}
.
\end{eqnarray}
For $m=0$ we have a system of integral equations that after the Fourier 
transformation gains the matrix form
\begin{equation}
H_{ij}(k)=C_{ij}(k)+\sum_{i'j'}C_{ii'}(k)\rho_{i'j'}H_{j'j}(k)
\label{15},
\end{equation}
where
$H_{ij}(k)=h_{ij0} (k)$,
$C_{ij}(k)=c_{ij0} (k)$,
$\rho_{ij}=\rho\left\langle Y_{i0} (\omega)Y_{j0}
(\omega)\right\rangle_{\omega}$, indices take the values 0 and 1. The 
problem of a self-consistent solution of the anisotropic OZ and Lovett 
equations has reduced at this stage to the solution of Eqs.~(\ref{14}) and 
(\ref{15}) under conditions (\ref{11}) and the self-consistent $f(\omega)$ 
given by relation (\ref{12}).  On the basis of the Wertheim-Baxter 
factorization method \cite{Wertheim} one can find the analytical solution 
of such equations in the form of a set of algebraic equations. The 
detailed derivation of similar solutions can be found in the literature, 
and therefore we omit any details and refer the reader to the previous 
publications \cite{PhysA,UFZH}. The explicit form of the 
solution is quite unwieldy and will be given in the extended 
version of this paper.  Here we point out only that it is efficiently 
computable, and we use it for calculation of isotherms by the virial route 
to thermodynamics in order to locate the gas-liquid transition by the 
Maxwell construction.

\begin{figure}
\noindent
\epsfxsize=57mm
\epsffile{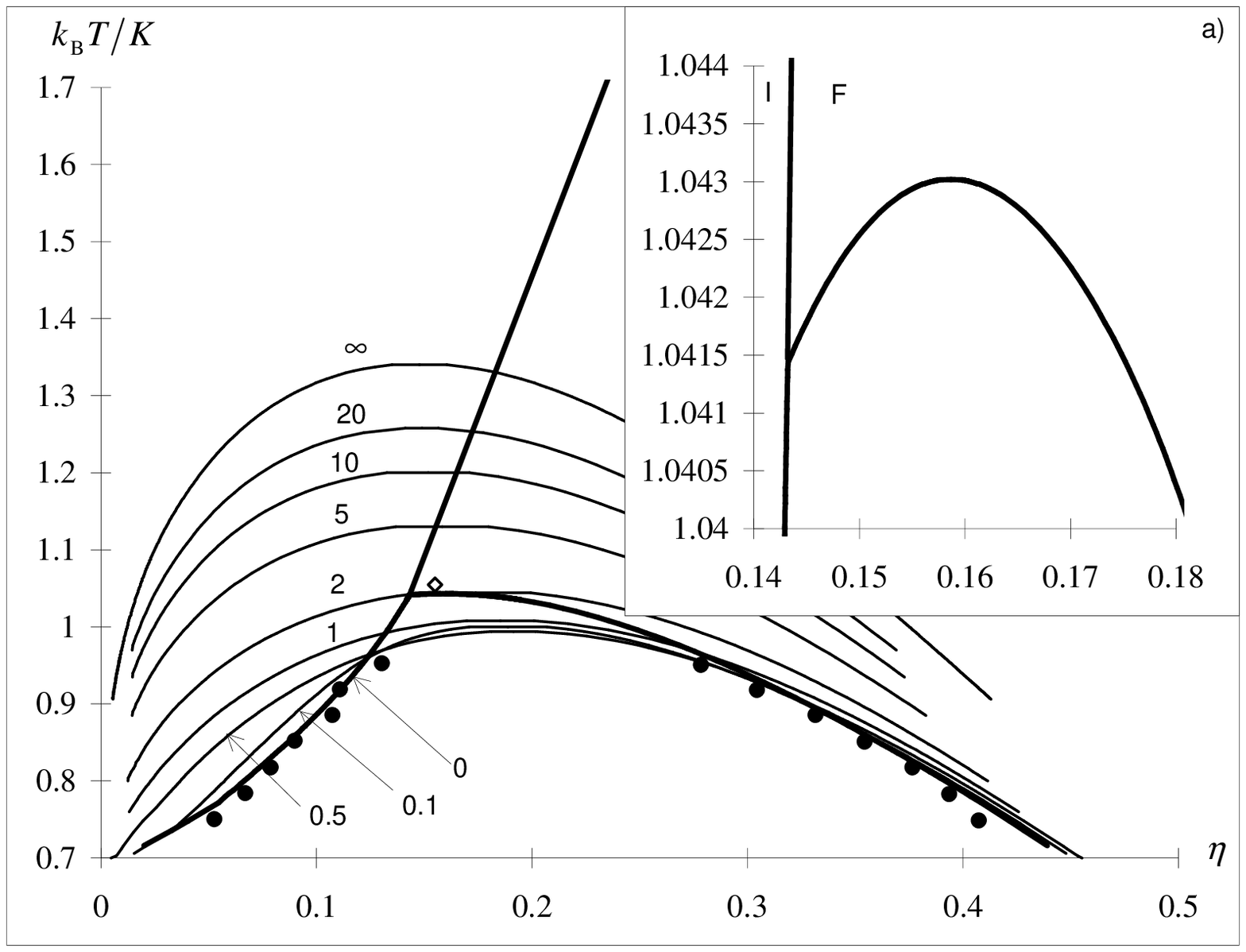}
\epsfxsize=57mm
\epsffile{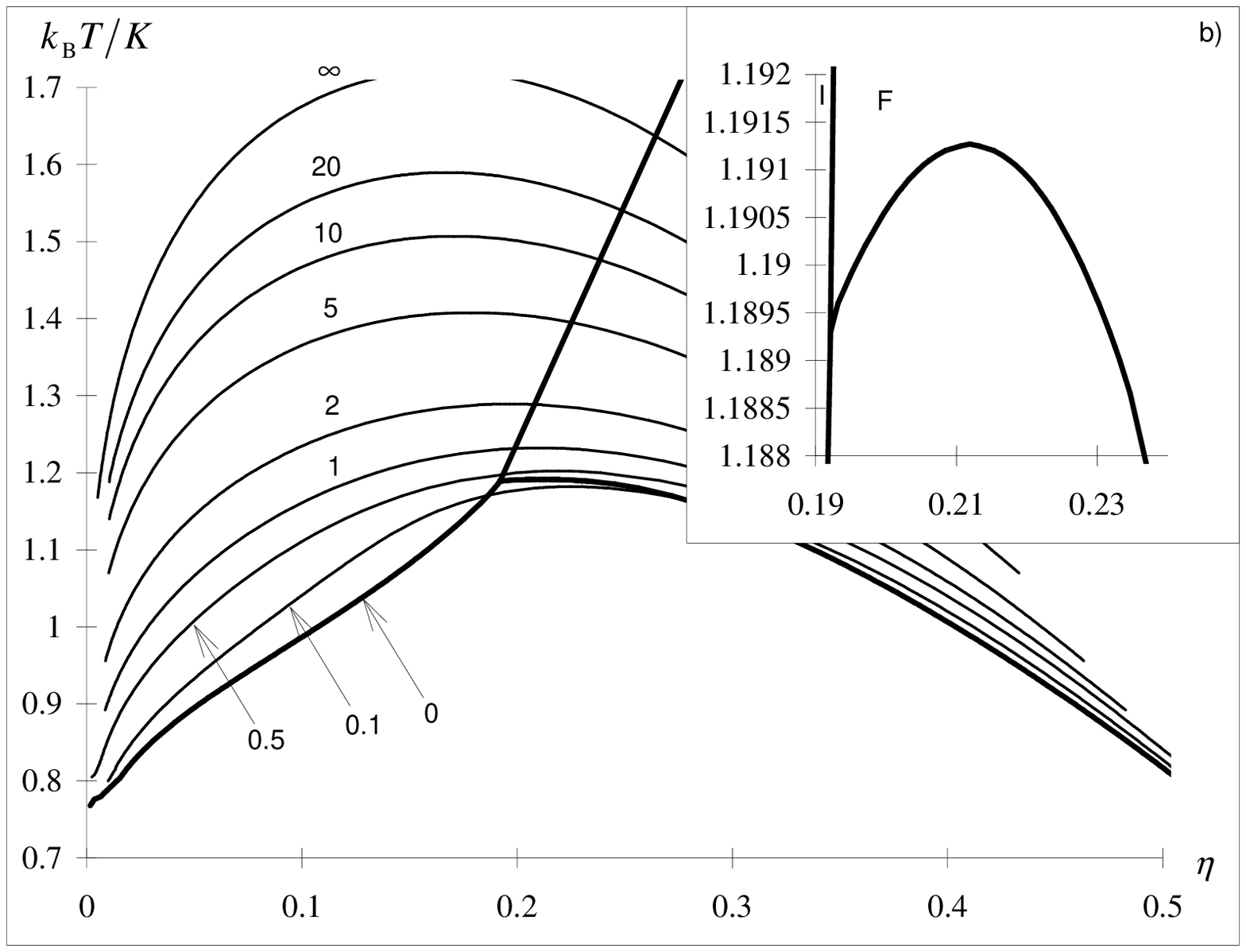}
\epsfxsize=57mm
\epsffile{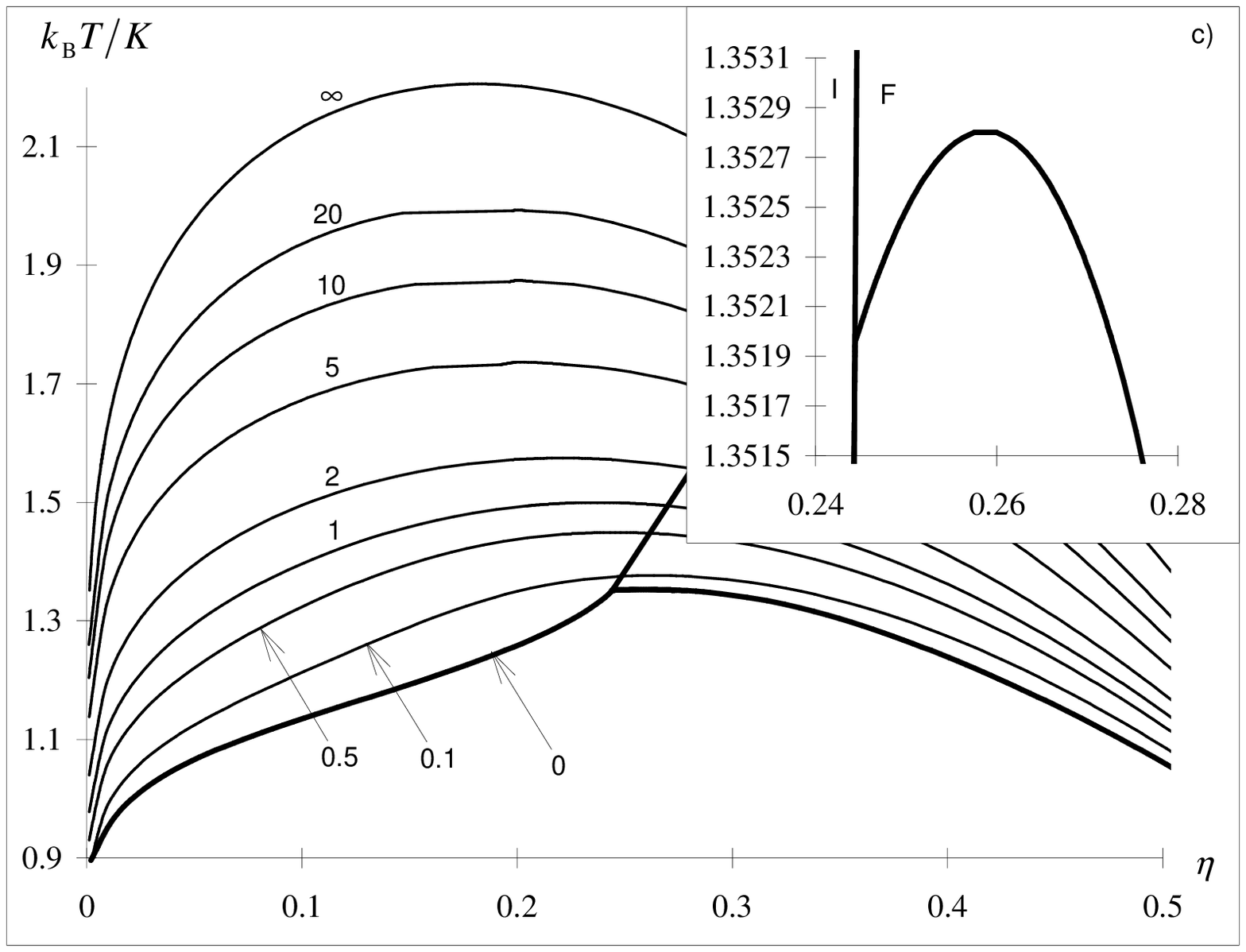}
\caption{
Phase diagram of the Heisenberg fluid in the external field for 
$z\sigma=1$ (a), $z\sigma=2$ (b), $z\sigma=3$ (c).  Lines are results of 
the MSA. The thick line is the case of zero 
field. On the insets the vicinity of the ferrogas-ferroliquid critical 
point is shown, I and F mark isotropic and ferromagnetic domains. The thin 
lines are the gas-liquid binodals of the fluid in the magnetic field (the 
attached numbers are the values of the field $h$). Simulation data 
\protect\cite{Lomba} ($h=0$) 
for $z\sigma=1$ (a) are shown as black circles (the 
gas-liquid coexistence) and diamonds (Curie points). It should be noted 
that $k_{\rm B}T/K$ equals $T^*/6$ from Ref.~\protect\cite{Lomba}.  
} 
\label{msa} 
\end{figure}

The configuration of the MSA zero-field phase diagrams slightly differs 
from that of the MF theory (Fig.~\ref{mfa}). The MSA via the virial route 
to thermodynamics demonstrates (Fig.~\ref{msa}) the lack of the 
tricritical point \cite{PhysA} in the Heisenberg fluid (\ref{1})--(\ref{2}) 
contrary to the MF and modified MF theories 
\cite{Tavares}.  Within the MSA the Curie line joins the gas-liquid 
binodal at its vapor branch (see insets to Fig.~\ref{msa}). 
This result is in whole agreement with the available zero-field simulations 
for the same model \cite{Lomba}: the liquid phase is ferromagnetic and the 
gaseous phase is mainly paramagnetic, except in the neighborhood of the 
critical point, where the transition ferroliquid-ferrogas takes place.  
The quantitative agreement with simulations is also quite perfect (see 
Fig.~\ref{msa}a). In the insets to the figure one can distinguish the 
critical point (CP, the top of the gas-liquid binodal) and 
the critical endpoint (CEP) in which the Curie line joins the binodal. In 
the temperature range from $t_{\rm CEP}$ to $t_{\rm CP}$ three spatially 
uniform phases (isotropic gas, ferrogas, ferroliquid) can exist. 
For long-range enough potentials this interval
($t_{\rm CP}-t_{\rm CEP}$) decreases with the decrease of $z\sigma$ (see 
table~\ref{t1}) and tends to zero for small values of $z\sigma$: the CP 
and the CEP coincide and form the tricritical point as is shown in 
Fig.~\ref{mfa}. In practice we can not distinguish the CP and the CEP by 
the MSA virial route already at $z\sigma=0.1$.

The effect of strong magnetic fields on the Heisenberg fluid 
(\ref{1})--(\ref{2}) consists in a considerable increase of the critical 
temperature. Strong external fields spread the gas-liquid coexistence 
region on the phase diagram, in other words, favors the gas-liquid phase 
separation. This result agrees with the conclusions of 
Refs.~\cite{VRP,Lado,LadoPRL}. But it follows from the MSA phase diagrams 
in Fig.~\ref{msa} that small fields can suppress the gas-liquid transition 
at finite values of $z\sigma$ (not only in the limit $z\sigma\to0$).  For 
example, one can see on Fig.~\ref{msa}a ($z\sigma=1$) that the external 
field of strength $h=0.1$ totally removes the phase separation at 
temperatures from $t_{\rm CP}(0.1)=1.002$ to $t_{\rm CP}(0)=1.043$.  In 
the systems with more short-range anisotropic interactions this 
temperature interval decreases.  For example, in Fig.~\ref{msa}b 
($z\sigma=2$) the temperature interval, where the external field $h=0.1$ 
removes the gas-liquid separation, is much less. For short-range enough 
potentials even very weak fields do not suppress separation, e.g., for 
$z\sigma=3$ the same field $h=0.1$ increases $t_{\rm CP}$ 
(Fig.~\ref{msa}c).  Thus, the MSA predicts that the suppression effect of 
small fields on the gas-liquid separation gets weak and finally disappears 
for short-range enough spin-spin interactions.  The verification of this 
conclusion we address to future investigations.  

\begin{table}
\begin{tabular}{lllll}
$z\sigma$&$t_{\rm CP}$&
$\eta_{\rm CP}$&
$t_{\rm CP}-t_{\rm CEP}$&
$\eta_{\rm CP}-\eta_{\rm CEP}$\\ 
\hline
2             & 1.1913 & 0.2119 & 0.0016 & 0.0185 \\
1             & 1.0430 & 0.1587 & 0.0015 & 0.0151 \\
0.5           & 0.9811 & 0.1316 & 0.0008 & 0.0063 \\
0 (MF theory) & 0.938  & 0.117  & 0      & 0  
\end{tabular}
\caption{Coordinates of the gas-liquid critical point and its 
distance from the critical endpoint at different $z\sigma$}
\label{t1}
\end{table}

{\em Acknowledgements.}
We are grateful to I.~M.~Mryglod for he compelled our attention to the 
problem. We thank J.~J.~Weis for sending us Ref.~\cite{Lado} prior to 
publication. We also thank H.~Iro and R.~Folk for giving us the paper
Ref.~\cite{Schinagl} prior to publication and for useful discussion.

\end{document}